\documentclass[twocolumn]{aastex631}

\usepackage{amsmath}
\usepackage{graphicx}
\usepackage{dcolumn}
\usepackage{bm}
\usepackage{xcolor}

\begin{document}

\title{Anisotropic particle acceleration in Alfv\'enic turbulence }

\correspondingauthor{Cristian Vega}
\email{csvega@wisc.edu}

\author{Cristian Vega}
\affiliation{Department of Physics, University of Wisconsin at Madison, Madison, Wisconsin 53706, USA}

\author[0000-0001-6252-5169]{Stanislav Boldyrev}
\affiliation{Department of Physics, University of Wisconsin at Madison, Madison, Wisconsin 53706, USA}
\affiliation{Center for Space Plasma Physics, Space Science Institute, Boulder, Colorado 80301, USA}

\author[0000-0003-1745-7587]{Vadim Roytershteyn}
\affiliation{Center for Space Plasma Physics, Space Science Institute, Boulder, Colorado 80301, USA}



\begin{abstract}
 Alfv\'enic turbulence is an effective mechanism for particle acceleration in strongly magnetized, relativistic plasma. In this study, we investigate a scenario where turbulent plasma is influenced by a strong guide magnetic field, resulting in highly anisotropic turbulent fluctuations. In such cases, the magnetic moments of particles are conserved, which means that acceleration can only occur along the direction of the magnetic field. Consistent with previous analytic studies, we find through PIC simulations of magnetically dominated pair plasma that the momenta of accelerated particles are closely aligned with the magnetic field lines. Notably, the alignment angle decreases as particle energy increases, potentially limited only by the inherent curvature and gradients of the turbulent magnetic fluctuations. This finding has significant implications for interpreting the synchrotron radiation emitted by highly accelerated particles.
\end{abstract}



\section{Introduction} \label{sec:intro}
In recent years, Alfvénic turbulence has been recognized as a significant mechanism for particle acceleration in collisionless plasmas. This acceleration is particularly efficient when the turbulence is magnetically dominated, meaning that the energy of the magnetic fluctuations exceeds the rest mass energy of the plasma particles. In such cases, the distribution function of the accelerated particles tends to be quite universal, depending only on the relative strength of the magnetic fluctuations compared to the guiding magnetic field \cite[e.g.,][]{zhdankin2017a,comisso2019,comisso2022,zhdankin2020,nattila2020,nattila2022,demidem2020,trotta2020,pezzi2022,vega2022a, vega2023,bresci2022}.

However, numerical simulations have revealed some intriguing results. When the guiding field is relatively weak in comparison with magnetic fluctuations (i.e., when $B_0 \lesssim \delta B_0$), the energy distribution function of the accelerated particles follows a power-law form \cite[e.g.,][]{zhdankin2017a,zhdankin2018,zhdankin2018c,comisso2019,comisso2022,vega2022a}. In contrast, when the guiding field is strong (i.e., when $B_0 \gg \delta B_0$), the distribution is better characterized by a lognormal form \cite[][]{vega2024,vega2024b}.

An explanation for this behavior was proposed in \cite[][]{vega2024b}. It suggests that the interaction between particles and turbulence is qualitatively different in two scenarios: weak and strong guide fields. In the case of a weak guide field, the particle gyroradii can be comparable to the sizes of turbulent eddies, resulting in a more efficient interaction with turbulence. Conversely, in a strong guide field, the particle gyroradius is always much smaller than the scales of the turbulent eddies. In this scenario, acceleration occurs through parallel electric fields and magnetic curvature drifts, which are less efficient due to the strong guide field.

This analysis indicates that, under a strong guide field, the particle's magnetic moment (an adiabatic invariant) is well conserved. Consequently, acceleration can only increase the particle's parallel momentum. As a result, the pitch angles of accelerated particles should become progressively smaller as the energy of the particles increases. This conclusion has significant implications for the interpretation of energetic cyclotron radiation produced by accelerated particles, as this type of radiation is highly dependent on the pitch angle \cite[e.g.,][]{zhdankin2020b,zhdankin2021,sobacchi2019,sobacchi2020,nattila2022,sobacchi2023,comisso2020,comisso2023,comisso2024}. {Numerical studies of the pitch angle evolution and corresponding synchrotron radiation in the case of a moderate guide field ($ B_0\sim \delta B_0 $) were conducted in \cite[][]{comisso2020}}.

{In this study, we perform particle-in-cell numerical simulations to explore turbulence and particle acceleration in a magnetically dominated pair plasma in the limit of a strong guide field. In this limit, the particle's magnetic moment is preserved throughout the acceleration process. We observe that the accelerated particles maintain small pitch angles in relation to the local magnetic field, and the pitch angle decreases as the particle energy increases. At high particle energies, the pitch angle distribution approaches a saturation point at which its scaling behavior with energy changes. We attribute this saturation to a modified form of the (still conserved) magnetic moment, which is influenced by weak magnetic curvature and gradient drifts associated with turbulent fluctuations, as predicted analytically in~\cite[][]{vega2024b}.}

\section{Numerical setup}

We performed 2.5D numerical simulations of decaying turbulence in a pair electron-positron plasma with fully relativistic particle-in-cell code VPIC \cite[][]{bowers2008}. A 2.5D simulation can be understood as follows. Given a 3D simulation with continuous translational symmetry along one direction, a single 2D cut perpendicular to this direction contains all the information of the 3D system. This cut would correspond to a 2.5D simulation. In our simulations, the two-dimensional domain was chosen perpendicular to the guide field. 

{This simplified 2.5D setup enables us to achieve a relatively high numerical resolution for studying turbulence at small kinetic scales. This approach also allows us to utilize a significantly larger number of particles per cell compared to fully 3D PIC simulations (see, for example, \cite[][]{nattila2022}). By preserving all vector components of both the electromagnetic field and particle momenta, this method is expected to capture essential nonlinear interactions that occur in three-dimensional scenarios. Previous numerical studies comparing 2.5D and 3D simulations have demonstrated that they produce similar energy spectra for both fields and particles \cite[e.g.,][]{zhdankin2017a, zhdankin2018c, comisso2018, comisso2019, vega2023}.}

The simulation domain was a double periodic $L\times L$ square and had a uniform magnetic guide field $\bm{B}_0=B_0\hat{\bm{z}}$. Turbulence was initialized with randomly phased magnetic perturbations of the shear-Alfv\'en type
\begin{eqnarray}
\delta{\bm B}(\mathbf{x})=\sum_{\mathbf{k}}\delta B_\mathbf{k}\hat{\xi}_\mathbf{k}\cos(\mathbf{k}\cdot\mathbf{x}+\phi_\mathbf{k}),
\end{eqnarray}
where the unit polarization vectors are normal to the background field, $\hat{\xi}_\mathbf{k}=\mathbf{k}\times {\bm B}_0/|\mathbf{k}\times{\bm B}_0|$. The wave vectors of the modes are given by $\mathbf{k}=\{2\pi n_x/L,{2}\pi n_y/L\}$, where $n_x,n_y=1,...,8$. All modes have the same amplitudes $\delta B_{\mathbf{k}}$. The initial root-mean-square value of the perturbations is given by $\delta B_0={\langle |\delta{\bm B}(\mathbf{x})|^2 \rangle^{1/2} }$, and the relative strength of the guide field is~$B_0/\delta B_0=10$. The outer scale of turbulence, $l=L/8$, and the speed of light, $c$, define the time scale $l/c$, which we used to normalize time in the results. 

The parameters of the runs are summarized in Table~\ref{table}. Run I had 50 particles per cell per species. To evaluate the potential for unphysical pitch-angle scattering caused by the minimal numerical noise that is inherently present in PIC simulations, the simulation was repeated in Run II with quadruple the number of particles per cell.

\begin{table}[t!]
\vskip5mm
\centering
\begin{tabular}{c c c c c c} 
\hline
{Run} & size $(d_e^2)$ & \# of cells & $\omega_{pe}\delta t$  & \# ppc \\
\hline
I & $1600^2$ & $23552^2$ & $6.0\times10^{-3}$ & 50 \\ 
II & $1600^2$ & $23552^2$ & $6.0\times10^{-3}$ & 200 \\ 
\hline
\end{tabular}
\caption{Parameters of the runs. Here, $d_e$ is the nonrelativistic electron inertial scale, $\omega_{pe}$ the nonrelativistic electron plasma frequency, and the last column lists the number of particles per cell in each run. In both runs, the initial ratio of the guide field to the perturbation is~$B_0/\delta B_0=10$, and the plasma magnetization is ${\sigma_0=4000}$. }
\label{table}
\end{table}

The plasma magnetization is defined as
\begin{eqnarray}
\sigma_0=\frac{B_0^2}{4\pi n_0 w_0 m_e c^2},    
\end{eqnarray}
where $n_0$ is the mean density of each species, and $w_0m_e c^2$ is the initial enthalpy per particle. To generate the initial distributions of both the electrons and the positrons, we used the isotropic Maxwell-J\"uttner function, with the temperature parameter $\Theta_0=k_BT_e/m_ec^2=0.1$.  For such a distribution, the specific enthalpy is given by $w_0=K_3(1/\Theta_0)/K_2(1/\Theta_0)\approx 1.27$, 
where $K_\nu $ is the modified Bessel function of the second kind. The magnetization in our simulations was $\sigma_0=4000$.

As in our previous studies \cite[e.g.,][]{vega2024}, we introduce an initial plasma current into the system to compensate for the curl of the initial magnetic perturbations, represented as $J_z=(c/4\pi){\bm \nabla}\times {\delta \bm B}_{\perp,0}$. This approach helps reduce the fraction of high-frequency ordinary modes that may be generated by decaying initial perturbation in addition to the Alfv\'en modes. 

To incorporate the current, we apply a velocity $U_{z}^s = J_z/(2 q_s n_0)$ (in a Newtonian fashion) to each particle of species $s$, where $q_s = \pm |e|$ (representing positrons and electrons) is sampled from the Maxwell-Jüttner distribution. This adjustment is made under the condition that $|{\bm v}^s + {\bm U}^s| < 0.97 c$, to avoid an artificial generation of very energetic particles. In regions where such an increase in velocity would lead to $|{\bm v}^s + {\bm U}^s| \geq  0.97 c$, the distribution remains unchanged. The addition of this current does not alter the core of the particle energy distribution function but generates a weak tail extending up to~$\gamma\approx 4$, as seen in Figure~\ref{particle_energy}.

\section{Results}

\begin{figure}[t!]
\centering
\includegraphics[width=\columnwidth]{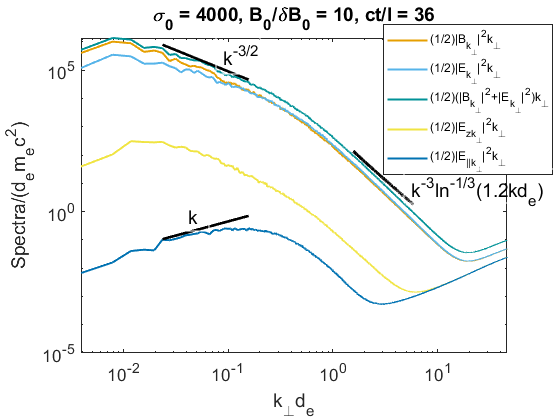}
\includegraphics[width=\columnwidth]{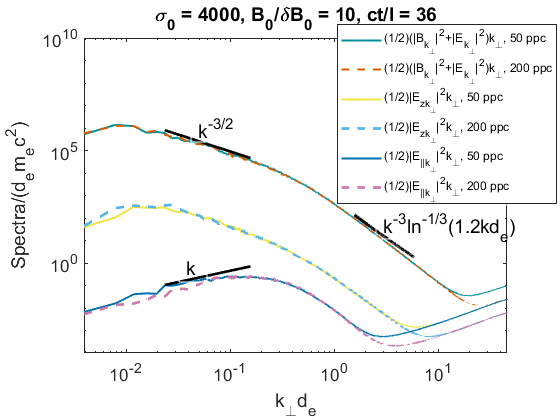}
\caption{Spectra of magnetic and electric fluctuations in the two runs. The top panel displays the results from Run~I. The bottom panel shows the results from Run~I (solid lines) overlaid with the results from Run~II (dashed lines) for comparison. Here, the subscript~$\perp$ denoted the field components perpendicular to the $z$-direction. The parallel component of the electric field is defined as a projection on the direction of the local magnetic field. The rising parts of the curves observed at large wave numbers are due to the numerical noise existing in the PIC simulations. The noise level is reduced in Run~II.  The slopes indicated by solid lines are given for the reader's orientation; their discussion can be found in~\cite[][]{vega2024}.
\label{spectra}}
\end{figure}

Figure~\ref{spectra} shows the Fourier energy spectra of the magnetic and electric fields in the two runs. All the runs have similar initial setups, however, Run~II has four times as large the number of particles per cell. 
{In this work, we focus on the limit of a strong guide field. In this case, the scales associated with the typical initial gyroscales of the particles, described by the equation $\rho_0 = {m_e c^2}/{eB_0}$, {are comparable to the cell size and} to the scales at which numerical noise dominates the spectra of magnetic and electric turbulent fluctuations.}

{Since the noise is weaker than the energy-containing magnetic fluctuations and is not expected to have frequencies comparable to the particles' gyrofrequencies, it should not impact the conservation of the electron magnetic moments. However, our study concentrates on particles accelerated along the magnetic field lines. These particles have extremely small pitch angles, so even minimal numerical pitch-angle scattering may affect the results. To evaluate the effects of numerical noise, we will, therefore, compare the results obtained in runs I and II, which differ only in the number of particles per cell.}

\begin{figure}[t!]
\centering
\includegraphics[width=\columnwidth]{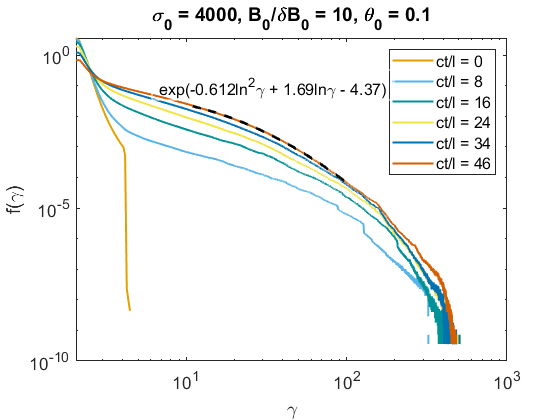}
\caption{Time evolution of the particle energy distribution function in Run~I. At large times ($ct/l\gtrsim 34$), the non-thermal tail of the distribution function approaches the log-normal form (denoted by the dashed line). 
\label{particle_energy}}
\end{figure}

Figure~\ref{particle_energy} shows the time evolution of the electron energy distribution function. Since the energy is given by $\gamma m_ec^2$, where $\gamma$ is the relativistic Lorentz gamma factor, the energy distribution can be characterized by the distribution of the gamma factor. The distribution function approaches the log-normal shape at large time, which is consistent with the results previously obtained in \cite[][]{vega2024,vega2024b}. Our numerical analysis will be conducted for the time associated with nearly saturated energy distribution functions, $ct/l=36$. {At this point, the energy of the electromagnetic fluctuations has decreased by about half, and the average energy of the plasma particles energized by the turbulence is $\langle\gamma \rangle\approx 6$.} We will be interested in the non-thermally accelerated particles, with $\gamma\geq 20$.

In order to describe the measurements of the electron pitch angles, we note that the pitch angle is not a relativistic invariant; it depends on the reference frame where the measurement is performed. It is important to stress that the particle magnetic moment is not conserved in the laboratory frame. Rather, {in the limit of a strong guide field}, gyro-averaged magnetic moment remains an adiabatic invariant in the so-called ``E-cross-B'' frame, that is, the reference frame moving with the velocity ${\bm u}_{E}=c {\bm E}\times{\bm B}/B^2$ \cite[e.g.,][]{northrop1963,littlejohn1983,littlejohn1984}. 

\begin{figure}[ht!]
\centering
\includegraphics[width=\columnwidth]{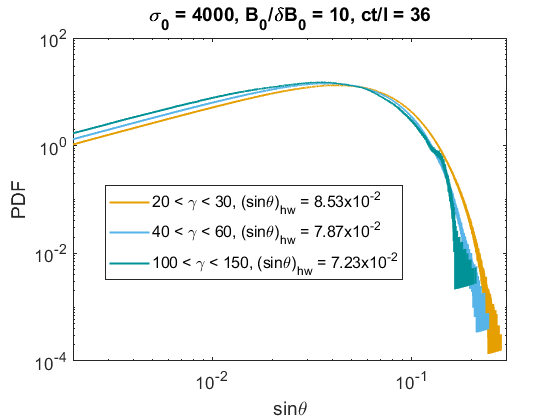}
\caption{The angular distribution of accelerated particles in the laboratory frame in Run~I. The half-width of the distribution functions is defined as the angle at which the magnitude of the distribution function decreases to half its maximum value. Notably, the measured half-width is independent of the particle energy and is primarily influenced by broadening effects due to the bulk $E\times B$ fluctuations in the plasma.
\label{angle_lab}}
\end{figure}

\begin{figure}[t!]
\centering
\includegraphics[width=\columnwidth]{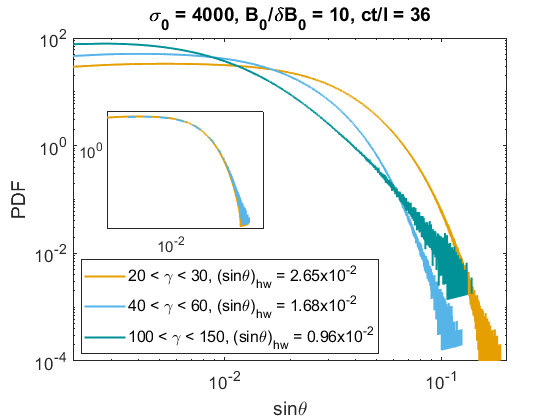}
\includegraphics[width=\columnwidth]{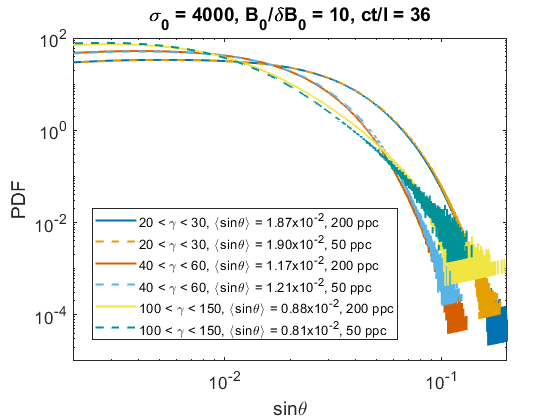}
\caption{The angular distributions of accelerated particles measured in the ``$E\times B$'' frame. The top panel shows the results of Run~I.   The measured half-widths of the probability density functions are presented. The inset shows that the orange curve with the argument rescaled according to 
$\sin \theta\to (\sin\theta)/1.6$ overlaps with the light blue curve, illustrating the self-similarity of the angular distribution function. {The bottom panel shows the results of Run~II (solid lines) overplotted with the results of Run~I (dashed lines). It also shows the values of $\langle \sin\theta\rangle$ calculated for each distribution function.}
\label{angle_EB}}
\end{figure}

We, therefore, need to boost both the particle momentum and the magnetic field measured in each cell to the corresponding ``E-cross-B'' frame. 
The results are illustrated in Figures~\ref{angle_lab} and \ref{angle_EB}. In Figure~\ref{angle_lab}, the pitch angles are measured in the laboratory frame. Since in magnetically dominated Alfv\'enic turbulence, $\sigma_0\gg 1$, the electric and magnetic fluctuations are on the same order, $E_\perp\sim \delta B_\perp$, the distribution saturates at the angles comparable to $\sin\theta\sim u_E/c\sim \delta B_0/B_0\sim 0.1$ at all energies. This implies that the in the laboratory frame, the pitch angular broadening is simply dominated by the bulk fluctuations of the plasma.   

The pitch angle distribution is, however, different when measured in the ``E-cross-B'' frame, see Figure~\ref{angle_EB} (a similar measurement procedure was also used in \cite[][]{comisso2020,nattila2022}). The typical pitch angles are now smaller, and their distributions depend on the particle energy. When the magnetic moment is conserved, the particle field-perpendicular momentum ${ p}_\perp$ cannot increase. The particle can then be accelerated only along the background magnetic field, that is, its field-parallel momentum should increase proportionally to the energy, ${p}_\|\propto \gamma$. The resulting small pitch angle will then become inversely proportional to the Lorentz factor, $\theta\sim p_\perp/p_\|\sim 1/\gamma$. Our numerical results presented in Figure~\ref{angle_EB} are broadly consistent with this behavior.  

{By comparing the two panels in Figure~\ref{angle_EB}, we observe that the number of particles per cell in our simulations does not significantly affect the angular distribution functions, although it minimally influences their far tails.}


\begin{figure}[t!]
\centering
\includegraphics[width=\columnwidth]{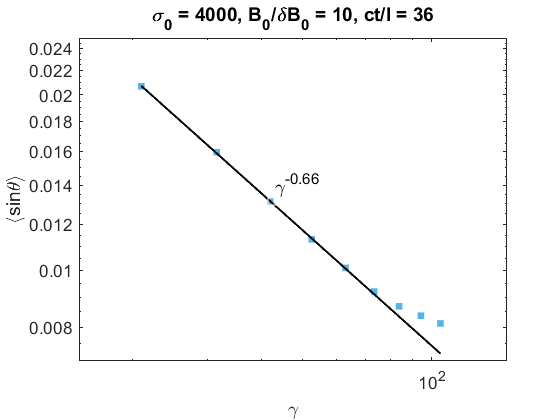}
\caption{The first moment, $\langle\sin\theta \rangle$, of the particle angular distribution functions, plotted as a function of energy. The scaling observed in Fig.~\ref{angle_EB} is illustrated by the solid line. The deviation from the indicated scaling occurs at energies of $\gamma\sim 100$, where the shape of the distribution function no longer follows self-similar behavior in Fig~\ref{angle_EB}. 
\label{sin_angle}}
\end{figure}

\section{Discussion}
We conducted a numerical study on particle acceleration caused by Alfvénic turbulence in a magnetically dominated pair plasma with ultrarelativistic particle temperatures. In the presence of a strong guide field, it is expected that the magnetic moment of the particles remains conserved. Consequently, the perpendicular momentum of the particles does not change, which means they can only be accelerated along the magnetic field lines. In this process, the gyroradius of the particles remains constant, leading to a decrease in the pitch angle as their energy increases. 

Our numerical results presented in Fig.~\ref{angle_EB} generally support this scenario. The rate at which the angular width decreases with energy is, however, slightly slower than predicted analytically. We observed that the typical angle decreases by a factor of approximately~1.6 when the energy doubles. The angular distribution function corresponding to this acceleration process has a self-similar form, as is evident from the inset in Figure~\ref{angle_EB}. Fig.~\ref{sin_angle} illustrates the scaling of the corresponding moments,~$\langle \sin\theta \rangle$. At higher energies, exceeding $\gamma \sim 100$ in our simulations, the collimation angle decreases with the energy at a slower rate, and the self-similar behavior of the distribution function is disrupted.

{The number of particles per cell has a minimal impact on the far tails of the distribution functions. 
To characterize the angular distributions shown in Fig. \ref{angle_EB}, we evaluate the half-maximum widths of the corresponding probability density functions. This width refers to the point where the function drops to half the value of its maximum.}

The high-energy behavior observed in Figs.~\ref{angle_EB}~\&~\ref{sin_angle}  may be related to the pitch-angle saturation effect, the physical phenomenon previously discussed in \cite[][]{vega2024b}. Even in the frame where the instantaneous “E-cross-B” drift is removed, particles still experience weak drifts due to magnetic line curvature, magnetic field gradients, and polarization drift caused by variations in the electric field along the particle's trajectory. For the magnetic and electric fluctuations generated by Alfv\'enic turbulence, these drifts are necessarily of the same order. We collectively refer to these as magnetic curvature effects, in contrast to the ``E-cross-B" drift, which can exist even in a uniform magnetic field.

In the presence of magnetic field variations, the standard magnetic moment, given by $\mu_0 = {p_\perp^2}/{2m_0B({\bm x})}$, is not conserved \cite[e.g.,][]{littlejohn1983,littlejohn1984,egedal2008}. In this expression, instead of assuming gyroaveraging, one considers the instantaneous values of the particle momentum ${\bm p}_\perp(t)$ and position ${\bm x}(t)$. The conserved quantity, instead, takes the form $\mu = \mu_0 + \mu_1 + \mu_2 + \dots $, where the first-order term behaves as ${\mu_1}/{\mu_0} \sim \left({p_\|}/{p_\perp}\right)\left({\rho}/{R_c}\right)$. Here, $\rho = {c}/{\Omega_e} = {\gamma m_ec^2}/{|e|B}$ represents the formal particle gyroscale, and $R_c $ is the curvature radius of the corresponding magnetic field line. It is assumed that $\rho/R_c\ll 1$.

As a particle propagates in the field-parallel direction, the variations in the field-perpendicular momentum along its trajectory cannot be smaller than $\Delta p_\perp \sim {p_\| \rho}/{R_c}$. Consequently, the corresponding pitch-angle variations cannot be smaller than $\Delta \theta \sim {\rho}/{R_c}$. Given that the largest field-line curvature in Alfv\'enic turbulence occurs at the relativistic electron inertial scale $d_{rel}$,\footnote{This estimate assumes that the smallest eddies produced by Alfvénic turbulence have a scale of $d_{rel}$. However, a refined analysis \cite[][]{demidov2025,boldyrev2025} indicates that the Alfvénic turbulent cascade can be terminated at a larger scale due to tearing instability or charge starvation effects. In these cases, the smallest Alfv\'enic scale, $d_{rel}$, should be replaced with the corresponding tearing scale or charge starvation scale.} we can estimate the field-line curvature as \cite[e.g.,][]{vega2024b}:
\begin{eqnarray}
\label{Rc}
R_c\sim {l} \left( \frac{B_0}{\delta B_0} \right)^2 \left(\frac{d_{rel}}{l}\right)^{1/3}.   
\end{eqnarray}

{{Parenthetically, we note that formula (\ref{Rc}) provides an order-of-magnitude estimate for the curvature radius. A more detailed analysis leads to the expression 
\begin{eqnarray}
R_c\sim \frac{l}{g} \left( \frac{B_0}{\delta B_0} \right)^2 \left(\frac{d_{rel}}{l}\right)^{1/3}, 
\end{eqnarray}
where $g$ is a numerical factor that accounts for the geometry and statistical distribution of the turbulent structures. For the structures examined in, for example, \cite[][]{vega2024b}, one estimates $g=2$. Additionally, the presence of strong and intermittent magnetic fluctuations $\delta B_0$ could result in smaller curvature radii and larger factors~$g$.}}

In Equation~(\ref{Rc}), for simplicity, we used the \citet{goldreich_toward_1995} model for the scaling of turbulent fluctuations. If instead one used the model of scale-dependent dynamic alignment~\cite[e.g.,][]{boldyrev2006,mason2006,mason2012,boldyrev2009,chandran_intermittency_2015,chen2016,walker2018,kasper2021,chernoglazov2021}, one would get a slightly different scaling for the curvature radius:
\begin{eqnarray}
\label{Rc2}
R_c\sim {l} \left( \frac{B_0}{\delta B_0} \right)^2 \left(\frac{d_{rel}}{l}\right)^{1/4}.   
\end{eqnarray}

For the scaling given by~Eq.~(\ref{Rc}), we estimate that for our runs, the intrinsic variations of the particle's pitch angle are on the order
\begin{eqnarray}
\label{rho_perp_small}
\sin\theta \sim \gamma\left(\frac{\rho_0}{l} \right) \left( \frac{\delta B_0}{B_0} \right)^2 \left(\frac{l}{d_{rel}} \right)^{1/3},
\end{eqnarray}
where $\rho_0=m_ec^2/|e|B$. 
For particles with $\gamma \gtrsim 100$ we estimate $\sin\theta \sim 10^{-3}$. This value is smaller than the angular broadening $\sin \theta \sim 10^{-2}$, where angular saturation effects become noticeable in Figure~\ref{sin_angle}. This could indicate the previously mentioned point that formulae~(\ref{Rc}) and~(\ref{rho_perp_small}) may overlook a numerical factor related to geometric effects, intermittency, as well as the comparable contributions from polarization and gradient drifts. These factors all contribute to an increase in the intrinsic variations of the pitch angle.

\begin{figure}[t!]
\hskip-5mm\includegraphics[width=1.1\columnwidth]{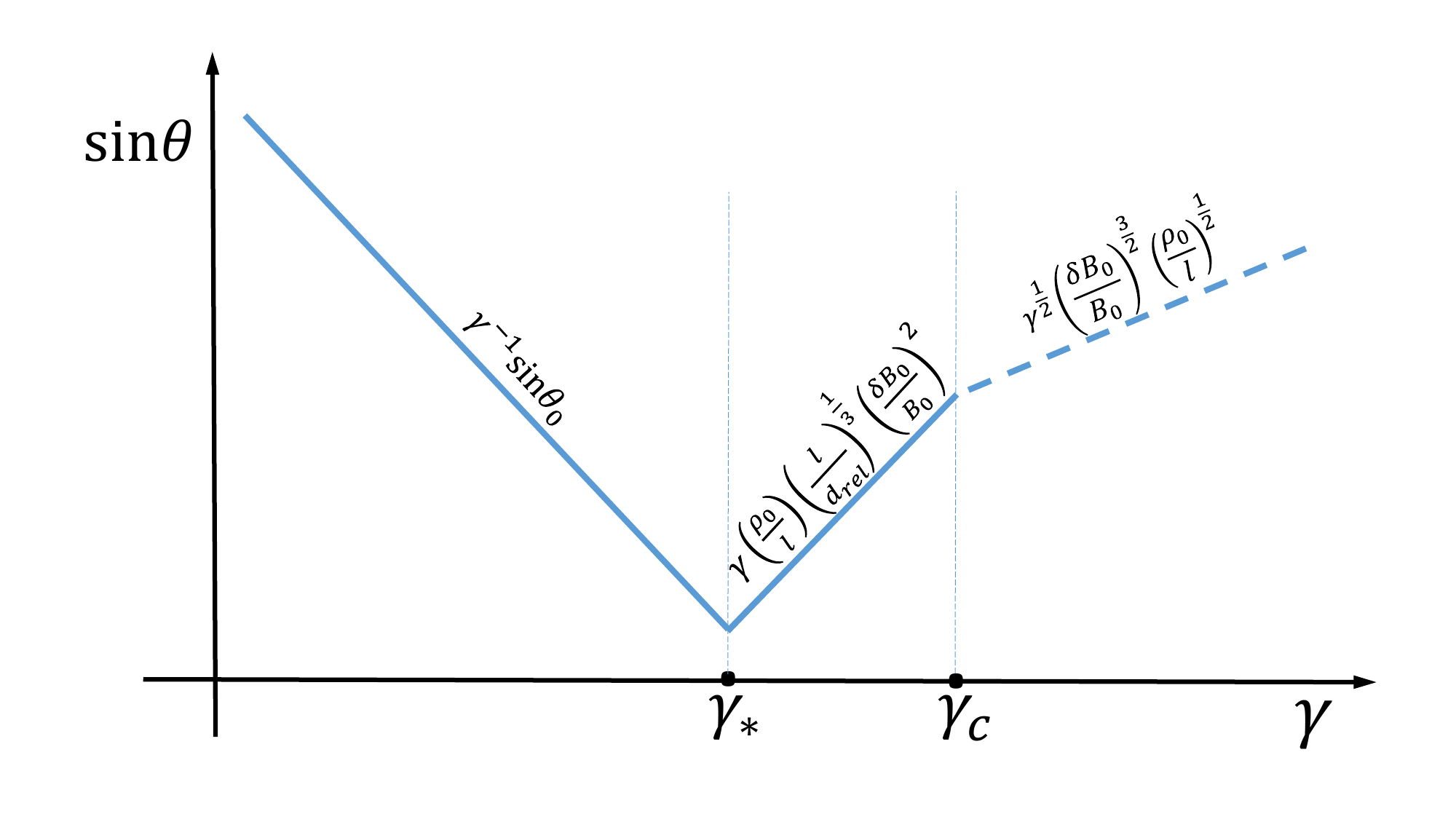}
\caption{Sketch on a log-log scale of the analytically modeled pitch angle of an ultrarelativistic electron accelerated by Alfv\'enic turbulence with a strong guiding field, as predicted by Eqs.~(\ref{sin_g}), (\ref{rho_perp_small}), and~(\ref{large_g}). {The dashed line indicates the region where the magnetic moment is not conserved due to the particle's interaction with turbulence.}}
\label{angles}
\end{figure}

\begin{figure}[t!]
\hskip-5mm\includegraphics[width=1.1\columnwidth]{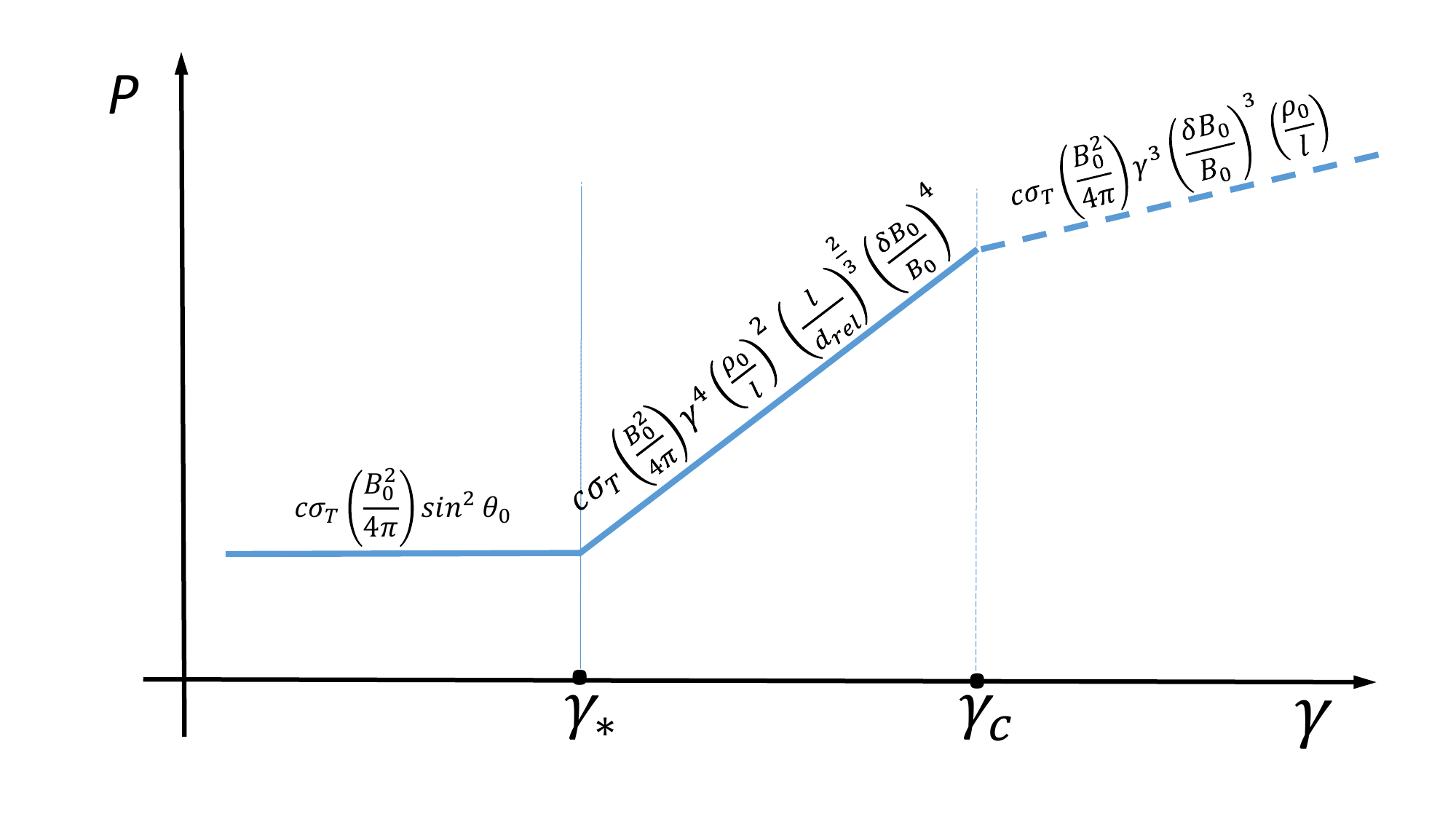}
\caption{Sketch on a log-log scale of the analytically modeled synchrotron radiation power generated by an ultrarelativistic electron accelerated by Alfv\'enic turbulence in the presence of a strong guide field, as predicted by Eqs. (\ref{p1}), (\ref{p2}), and (\ref{p3}). {The dashed line indicates the region where the magnetic moment is not conserved due to the particle's interaction with turbulence.}}
\label{power}
\end{figure}

Finally, we briefly discuss synchrotron radiation produced by ultrarelativistic electrons energized by turbulence with a strong guide field. Assume that the typical initial pitch angle of a particle is $\sin\theta_0\sim 1$. As it gets accelerated by turbulence with a strong guide field, its magnetic moment is conserved and the typical pitch angle decreases according to 
\begin{eqnarray}
\sin \theta\sim \gamma^{-1}\sin\theta_0.
\label{sin_g}
\end{eqnarray}
The power of its synchrotron radiation averaged over the angular distribution is then independent of its energy,
\begin{eqnarray}
\label{p1}
P\sim c\sigma_T\frac{B_0^2}{4\pi}\gamma^2\sin^2\theta\sim c\sigma_T\frac{B_0^2}{4\pi}\sin^2\theta_0,
\end{eqnarray}
where $\sigma_T$ is the Thomson electron cross section. This expression, however, holds until the pitch angle decreases to the value given by Eq.~(\ref{rho_perp_small}). This happens when the energy reaches the characteristic value\footnote{{Based on the parameters of the runs, one estimates $\gamma_* \sim 500$. However, the order-of-magnitude estimate given by Eq.~(\ref{gamma_*}) does not account for numerical factors arising from geometrical and intermittency effects, which lead to a decrease in $\gamma_*$, potentially bringing it more in line with the numerically observed value of $\gamma_* \gtrsim 100$.}}
\begin{eqnarray}
\gamma_*\sim {\sin^{1/2}\theta_0}\left(\frac{l}{\rho_0}\right)^{1/2}\left(\frac{d_{rel}}{l}\right)^{1/6}\frac{B_0}{\delta B_0}. 
\label{gamma_*}
\end{eqnarray}
At higher energies, $\gamma>\gamma_*$, the pitch angle increases with the energy according to Eq.~(\ref{rho_perp_small}), and the pitch-angle averaged radiation power rapidly increases with the Lorentz factor:
\begin{eqnarray}
\label{p2}
P\sim c\sigma_T\frac{B_0^2}{4\pi}\gamma^4  \left(\frac{\rho_0}{l}\right)^{2}\left(\frac{l}{d_{rel}}\right)^{2/3}\left(\frac{\delta B_0}{B_0}\right)^4.
\end{eqnarray}

At even higher energies, $\gamma>\gamma_c$, where 
\begin{eqnarray}
\gamma_c=\frac{B_0}{\delta B_0}\frac{l}{\rho_0}\left(\frac{d_{rel}}{l}\right)^{2/3},
\label{gamma_c}
\end{eqnarray}
the particle's gyroradus becomes comparable to the electron inertial scale $d_{rel}$, the smallest scale associated with the Alfv\'enic fluctuations. {At such energies, a particle can interact more efficiently with turbulent fluctuations, and its magnetic moment can change due to pitch-angle scattering. As proposed in \cite[][]{vega2024b}, in this regime the particle's pitch angle scales with energy as}:
\begin{eqnarray}
\sin\theta \sim\gamma^{1/2}\left(\frac{\delta B_0}{B_0} \right)^{3/2}\left(\frac{\rho_0}{l} \right)^{1/2}.   
\label{large_g}
\end{eqnarray}
In this case, the power of synchrotron radiation changes to
\begin{eqnarray}
\label{p3}
P\sim c\sigma_T\frac{B_0^2}{4\pi}\gamma^3\left(\frac{\delta B_0}{B_0} \right)^3\frac{\rho_0}{l}.    
\end{eqnarray}
Figures~\ref{angles} and~\ref{power} summarize the scaling results presented by Equations~(\ref{sin_g}), (\ref{rho_perp_small}), and~(\ref{large_g}) and Equations~(\ref{p1}), (\ref{p2}),~and~(\ref{p3}).

\section{Conclusions}
{
Non-thermal particle acceleration is a fundamental problem in plasma astrophysics. Numerical simulations suggest that strong, magnetically dominated turbulence may serve as an efficient mechanism for particle acceleration; however, a complete understanding of this process is still lacking. Analytical studies indicate that the behavior of particles' magnetic moments may hold the key to understanding the acceleration process.} 

{We argue that the pitch angles of accelerated particles, which influence the radiation signatures of the relativistic plasma \cite[e.g.,][]{sobacchi2019,sobacchi2020,sobacchi2023,comisso2020,nattila2022,comisso2023,comisso2024}, depends critically on the conservation of particles' magnetic moments. Furthermore, analytical modeling of turbulent particle acceleration \cite[e.g.,][]{lemoine2021,vega2024b} significantly depends on whether the magnetic moments are preserved during the acceleration process or 
disrupted by interactions with turbulent fluctuations.}

{Our study has focused on magnetically dominated turbulence in the presence of a strong guide field. In this limit, the magnetic moments are expected to be well conserved throughout the acceleration process.  Our numerical results support the conservation of the magnetic moment, indicating that the pitch angle decreases with increasing energy in a self-similar fashion up to the point defined by $\gamma_*$ in Equation ~(\ref{gamma_*}). At higher energies, the pitch-angle scaling behavior is expected to change due to a more complex form of the (still conserved) magnetic moment. This continues until very high energy, $\gamma_c$,  is reached, as shown in Fig.~\ref{angles}, at which point the conservation of the magnetic moment is broken.}  

{Given that $\gamma_c \sim \sqrt{d_{rel}/\rho_0}\,\gamma_*$, the separation of these energy scales is significant in the limit of a strong guide field. In this context, our study complements the work by \cite[][]{comisso2020}, which numerically examined turbulence at moderate guide fields, $B_0/\delta B_0 \leq 2$. Our analytical prediction, illustrated in Fig.~\ref{angles}, however, qualitatively agrees with their numerical observations of the pitch-angle scaling at the largest extreme of $B_0/\delta B_0 = 2$. Our results also qualitatively agree with the fully 3D numerical studies conducted by \cite[][]{nattila2022}, which observed a weaker decline in the pitch angle with increasing energy. However, these simulations used lower numerical resolution and significantly fewer particles per cell compared to our 2.5D studies, which may complicate the measurements of very small pitch angles.} 

{We believe that our results will be useful for developing analytical models of particle acceleration in relativistic turbulent plasmas as well as for interpreting numerical and observational data.}


\begin{acknowledgments}
Stimulating conversations with Yuri Lyubarsky, Mikhail Medvedev, and Alexander Philippov are gratefully acknowledged. 
We thank the anonymous reviewer for the constructive comments that allowed us to improve the text. 
This work was supported by the U.S. Department of Energy, Office of Science, Office of Fusion Energy Sciences under award number DE-SC0024362. VR was also partly supported by NASA grant 80NSSC21K1692. Computational resources were provided by the Texas Advanced Computing  Center (TACC) at the University of Texas at Austin and by the NASA High-End Computing (HEC) Program through the NASA Advanced Supercomputing (NAS) Division at Ames Research Center. 
This research also used resources of the National Energy Research Scientific Computing Center, a DOE Office of Science User Facility supported by the Office of Science of the U.S. Department of Energy under Contract No. DE-AC02-05CH11231 using NERSC awards FES-ERCAP0028833 and FES-ERCAP0033257. 
\end{acknowledgments}


\bibliography{references}{}

\begin{thebibliography}{}
\expandafter\ifx\csname natexlab\endcsname\relax\def\natexlab#1{#1}\fi
\providecommand{\url}[1]{\href{#1}{#1}}
\providecommand{\dodoi}[1]{doi:~\href{http://doi.org/#1}{\nolinkurl{#1}}}
\providecommand{\doeprint}[1]{\href{http://ascl.net/#1}{\nolinkurl{http://ascl.net/#1}}}
\providecommand{\doarXiv}[1]{\href{https://arxiv.org/abs/#1}{\nolinkurl{https://arxiv.org/abs/#1}}}

\bibitem[{{Boldyrev}(2006)}]{boldyrev2006}
{Boldyrev}, S. 2006, \prl, 96, 115002, \dodoi{10.1103/PhysRevLett.96.115002}

\bibitem[{{Boldyrev} \& {Loureiro}(2025)}]{boldyrev2025}
{Boldyrev}, S., \& {Loureiro}, N.~F. 2025, \apj, 979, 232, \dodoi{10.3847/1538-4357/ada28a}

\bibitem[{{Boldyrev} {et~al.}(2009){Boldyrev}, {Mason}, \& {Cattaneo}}]{boldyrev2009}
{Boldyrev}, S., {Mason}, J., \& {Cattaneo}, F. 2009, The Astrophysical Journal Letters, 699, L39, \dodoi{10.1088/0004-637X/699/1/L39}

\bibitem[{{Bowers} {et~al.}(2008){Bowers}, {Albright}, {Yin}, {Bergen}, \& {Kwan}}]{bowers2008}
{Bowers}, K.~J., {Albright}, B.~J., {Yin}, L., {Bergen}, B., \& {Kwan}, T.~J.~T. 2008, Physics of Plasmas, 15, 055703, \dodoi{10.1063/1.2840133}

\bibitem[{{Bresci} {et~al.}(2022){Bresci}, {Lemoine}, {Gremillet}, {Comisso}, {Sironi}, \& {Demidem}}]{bresci2022}
{Bresci}, V., {Lemoine}, M., {Gremillet}, L., {et~al.} 2022, \prd, 106, 023028, \dodoi{10.1103/PhysRevD.106.023028}

\bibitem[{{Chandran} {et~al.}(2015){Chandran}, {Schekochihin}, \& {Mallet}}]{chandran_intermittency_2015}
{Chandran}, B.~D.~G., {Schekochihin}, A.~A., \& {Mallet}, A. 2015, The Astrophysical Journal, 807, 39, \dodoi{10.1088/0004-637X/807/1/39}

\bibitem[{{Chen}(2016)}]{chen2016}
{Chen}, C.~H.~K. 2016, Journal of Plasma Physics, 82, 535820602, \dodoi{10.1017/S0022377816001124}

\bibitem[{{Chernoglazov} {et~al.}(2021){Chernoglazov}, {Ripperda}, \& {Philippov}}]{chernoglazov2021}
{Chernoglazov}, A., {Ripperda}, B., \& {Philippov}, A. 2021, \apjl, 923, L13, \dodoi{10.3847/2041-8213/ac3afa}

\bibitem[{{Comisso}(2024)}]{comisso2024}
{Comisso}, L. 2024, \apj, 972, 9, \dodoi{10.3847/1538-4357/ad51fe}

\bibitem[{{Comisso} \& {Jiang}(2023)}]{comisso2023}
{Comisso}, L., \& {Jiang}, B. 2023, \apj, 959, 137, \dodoi{10.3847/1538-4357/ad1241}

\bibitem[{{Comisso} \& {Sironi}(2018)}]{comisso2018}
{Comisso}, L., \& {Sironi}, L. 2018, \prl, 121, 255101, \dodoi{10.1103/PhysRevLett.121.255101}

\bibitem[{{Comisso} \& {Sironi}(2019)}]{comisso2019}
---. 2019, \apj, 886, 122, \dodoi{10.3847/1538-4357/ab4c33}

\bibitem[{{Comisso} \& {Sironi}(2022)}]{comisso2022}
---. 2022, \apjl, 936, L27, \dodoi{10.3847/2041-8213/ac8422}

\bibitem[{{Comisso} {et~al.}(2020){Comisso}, {Sobacchi}, \& {Sironi}}]{comisso2020}
{Comisso}, L., {Sobacchi}, E., \& {Sironi}, L. 2020, \apjl, 895, L40, \dodoi{10.3847/2041-8213/ab93dc}

\bibitem[{{Demidem} {et~al.}(2020){Demidem}, {Lemoine}, \& {Casse}}]{demidem2020}
{Demidem}, C., {Lemoine}, M., \& {Casse}, F. 2020, \prd, 102, 023003, \dodoi{10.1103/PhysRevD.102.023003}

\bibitem[{{Demidov} \& {Lyubarsky}(2025)}]{demidov2025}
{Demidov}, I., \& {Lyubarsky}, Y. 2025, \apj, 979, 104, \dodoi{10.3847/1538-4357/ad9d3a}

\bibitem[{{Egedal} {et~al.}(2008){Egedal}, {Fox}, {Katz}, {Porkolab}, {{\O}Ieroset}, {Lin}, {Daughton}, \& {Drake}}]{egedal2008}
{Egedal}, J., {Fox}, W., {Katz}, N., {et~al.} 2008, Journal of Geophysical Research (Space Physics), 113, A12207, \dodoi{10.1029/2008JA013520}

\bibitem[{{Goldreich} \& {Sridhar}(1995)}]{goldreich_toward_1995}
{Goldreich}, P., \& {Sridhar}, S. 1995, The Astrophysical Journal, 438, 763, \dodoi{10.1086/175121}

\bibitem[{Kasper {et~al.}(2021)Kasper, Klein, Lichko, Huang, Chen, Badman, Bonnell, Whittlesey, Livi, Larson, Pulupa, Rahmati, Stansby, Korreck, Stevens, Case, Bale, Maksimovic, Moncuquet, Goetz, Halekas, Malaspina, Raouafi, Szabo, MacDowall, Velli, Dudok~de Wit, \& Zank}]{kasper2021}
Kasper, J.~C., Klein, K.~G., Lichko, E., {et~al.} 2021, Phys. Rev. Lett., 127, 255101, \dodoi{10.1103/PhysRevLett.127.255101}

\bibitem[{{Lemoine}(2021)}]{lemoine2021}
{Lemoine}, M. 2021, \prd, 104, 063020, \dodoi{10.1103/PhysRevD.104.063020}

\bibitem[{{Littlejohn}(1983)}]{littlejohn1983}
{Littlejohn}, R.~G. 1983, Journal of Plasma Physics, 29, 111, \dodoi{10.1017/S002237780000060X}

\bibitem[{{Littlejohn}(1984)}]{littlejohn1984}
---. 1984, Physics of Fluids, 27, 976, \dodoi{10.1063/1.864688}

\bibitem[{{Mason} {et~al.}(2006){Mason}, {Cattaneo}, \& {Boldyrev}}]{mason2006}
{Mason}, J., {Cattaneo}, F., \& {Boldyrev}, S. 2006, \prl, 97, 255002, \dodoi{10.1103/PhysRevLett.97.255002}

\bibitem[{{Mason} {et~al.}(2012){Mason}, {Perez}, {Boldyrev}, \& {Cattaneo}}]{mason2012}
{Mason}, J., {Perez}, J.~C., {Boldyrev}, S., \& {Cattaneo}, F. 2012, Physics of Plasmas, 19, 055902, \dodoi{10.1063/1.3694123}

\bibitem[{{N{\"a}ttil{\"a}} \& {Beloborodov}(2021)}]{nattila2020}
{N{\"a}ttil{\"a}}, J., \& {Beloborodov}, A.~M. 2021, \apj, 921, 87, \dodoi{10.3847/1538-4357/ac1c76}

\bibitem[{{N{\"a}ttil{\"a}} \& {Beloborodov}(2022)}]{nattila2022}
---. 2022, \prl, 128, 075101, \dodoi{10.1103/PhysRevLett.128.075101}

\bibitem[{{Northrop}(1963)}]{northrop1963}
{Northrop}, Theodore, G. 1963, {The Adiabatic Motion of Charged Particles (Interscience Publishers, Inc., John Wiley \& Sons, New York)}

\bibitem[{{Pezzi} {et~al.}(2022){Pezzi}, {Blasi}, \& {Matthaeus}}]{pezzi2022}
{Pezzi}, O., {Blasi}, P., \& {Matthaeus}, W.~H. 2022, \apj, 928, 25, \dodoi{10.3847/1538-4357/ac5332}

\bibitem[{{Sobacchi} \& {Lyubarsky}(2019)}]{sobacchi2019}
{Sobacchi}, E., \& {Lyubarsky}, Y.~E. 2019, \mnras, 484, 1192, \dodoi{10.1093/mnras/stz044}

\bibitem[{{Sobacchi} \& {Lyubarsky}(2020)}]{sobacchi2020}
---. 2020, \mnras, 491, 3900, \dodoi{10.1093/mnras/stz3313}

\bibitem[{{Sobacchi} {et~al.}(2023){Sobacchi}, {Piran}, \& {Comisso}}]{sobacchi2023}
{Sobacchi}, E., {Piran}, T., \& {Comisso}, L. 2023, \apjl, 946, L51, \dodoi{10.3847/2041-8213/acc84d}

\bibitem[{{Trotta} {et~al.}(2020){Trotta}, {Franci}, {Burgess}, \& {Hellinger}}]{trotta2020}
{Trotta}, D., {Franci}, L., {Burgess}, D., \& {Hellinger}, P. 2020, \apj, 894, 136, \dodoi{10.3847/1538-4357/ab873c}

\bibitem[{{Vega} {et~al.}(2023){Vega}, {Boldyrev}, \& {Roytershteyn}}]{vega2023}
{Vega}, C., {Boldyrev}, S., \& {Roytershteyn}, V. 2023, \apj, 949, 98, \dodoi{10.3847/1538-4357/accd73}

\bibitem[{{Vega} {et~al.}(2024{\natexlab{a}}){Vega}, {Boldyrev}, \& {Roytershteyn}}]{vega2024}
---. 2024{\natexlab{a}}, \apj, 965, 27, \dodoi{10.3847/1538-4357/ad2e02}

\bibitem[{{Vega} {et~al.}(2024{\natexlab{b}}){Vega}, {Boldyrev}, \& {Roytershteyn}}]{vega2024b}
---. 2024{\natexlab{b}}, \apj, 971, 106, \dodoi{10.3847/1538-4357/ad5f8f}

\bibitem[{{Vega} {et~al.}(2022){Vega}, {Boldyrev}, {Roytershteyn}, \& {Medvedev}}]{vega2022a}
{Vega}, C., {Boldyrev}, S., {Roytershteyn}, V., \& {Medvedev}, M. 2022, \apjl, 924, L19, \dodoi{10.3847/2041-8213/ac441e}

\bibitem[{{Walker} {et~al.}(2018){Walker}, {Boldyrev}, \& {Loureiro}}]{walker2018}
{Walker}, J., {Boldyrev}, S., \& {Loureiro}, N.~F. 2018, Physical Review E, 98, 033209, \dodoi{10.1103/PhysRevE.98.033209}

\bibitem[{{Wong} {et~al.}(2020){Wong}, {Zhdankin}, {Uzdensky}, {Werner}, \& {Begelman}}]{zhdankin2020}
{Wong}, K., {Zhdankin}, V., {Uzdensky}, D.~A., {Werner}, G.~R., \& {Begelman}, M.~C. 2020, \apjl, 893, L7, \dodoi{10.3847/2041-8213/ab8122}

\bibitem[{{Zhdankin} {et~al.}(2021){Zhdankin}, {Uzdensky}, \& {Kunz}}]{zhdankin2021}
{Zhdankin}, V., {Uzdensky}, D.~A., \& {Kunz}, M.~W. 2021, \apj, 908, 71, \dodoi{10.3847/1538-4357/abcf31}

\bibitem[{{Zhdankin} {et~al.}(2018{\natexlab{a}}){Zhdankin}, {Uzdensky}, {Werner}, \& {Begelman}}]{zhdankin2018}
{Zhdankin}, V., {Uzdensky}, D.~A., {Werner}, G.~R., \& {Begelman}, M.~C. 2018{\natexlab{a}}, \apjl, 867, L18, \dodoi{10.3847/2041-8213/aae88c}

\bibitem[{{Zhdankin} {et~al.}(2018{\natexlab{b}}){Zhdankin}, {Uzdensky}, {Werner}, \& {Begelman}}]{zhdankin2018c}
---. 2018{\natexlab{b}}, \mnras, 474, 2514, \dodoi{10.1093/mnras/stx2883}

\bibitem[{{Zhdankin} {et~al.}(2020){Zhdankin}, {Uzdensky}, {Werner}, \& {Begelman}}]{zhdankin2020b}
---. 2020, \mnras, 493, 603, \dodoi{10.1093/mnras/staa284}

\bibitem[{{Zhdankin} {et~al.}(2017){Zhdankin}, {Werner}, {Uzdensky}, \& {Begelman}}]{zhdankin2017a}
{Zhdankin}, V., {Werner}, G.~R., {Uzdensky}, D.~A., \& {Begelman}, M.~C. 2017, \prl, 118, 055103, \dodoi{10.1103/PhysRevLett.118.055103}

\end{thebibliography}
\bibliographystyle{aasjournal}



\end{document}